\documentclass[aps,pre,preprint,showpacs,amsmath,amssymb,floatfix,tightenlines,showkeys]{revtex4-2} 

\usepackage[pdftex]{graphicx}       
\usepackage{xcolor}
\usepackage{txfonts}        
\usepackage[normalem]{ulem}

\newcommand{\rr}{{\mathbf r}}
\newcommand{\avc}{\langle m_2 \rangle / n}
\newcommand{\avd}{\langle m_3 \rangle / n}

\newcommand{\avR}{\langle R_n^2 \rangle}
\newcommand{\eref}[1]{Eq.~\eqref{#1}}
\newcommand{\fref}[1]{Fig.~\ref{#1}}
\newcommand{\sref}[1]{Section \ref{#1}}

\usepackage{hyperref} 
\hypersetup{colorlinks,citecolor=blue,filecolor=blue,linkcolor=blue,urlcolor=blue}

\begin{document}

\title{Polymer collapse of a self-avoiding trail model on a two-dimensional inhomogeneous lattice}

\author{C. J. Bradly} \email{chris.bradly@unimelb.edu.au}
\author{A. L. Owczarek}\email{owczarek@unimelb.edu.au}
\affiliation{School of Mathematics and Statistics, University of Melbourne, Victoria 3010, Australia}
\date{\today}

\begin{abstract}
The study of the effect of random impurities on the collapse of a flexible polymer in dilute solution has had recent attention with consideration of  semi-stiff  interacting self-avoiding walks on the square lattice.
In the absence of impurities the model displays two types of collapsed phase, one of which is both anisotropically ordered and maximally dense (crystal-like). In the presence of impurities the study showed that the crystal type phase disappears.  Here we investigate extended interacting self-avoiding trails on the triangular lattice with random impurities. Without impurities this model also displays two collapsed phases, one of which is maximally dense. However, this maximally dense phase is not ordered anisotropically. 
The trails are simulated using the flatPERM algorithm and the inhomogeneity is realised as a random fraction of the lattice that is unavailable to the trails.
We calculate several thermodynamic and metric quantities to map out the phase diagram and look at how the amount of disorder affects the properties of each phase but especially the maximally dense phase. Our results indicate that while the maximally dense phase in the trail model is affected less than in the walk model it is also disrupted and becomes a denser version of the globule phase so that the model with impurities only displays no more than one true thermodynamic collapsed phase.
\keywords{polymer collapse, inhomogeneous lattice, self-avoiding trails}
\end{abstract}

\maketitle

\section{Introduction}
\label{sec:Intro}

The effect of random disorder on otherwise well-understood statistical mechanical problems is an important topic, going back to the Ising model \cite{Watson1969}.
The study of the lattice polymer model on randomly diluted lattices goes back almost as far and has been closely related to the problem of percolation.
Fundamental scaling laws for the self-avoiding walk model of polymers persist on inhomogeneous lattices, provided the disorder is above the percolation limit $p_c$ \cite{Kremer1981, Duplantier1988}.
Change in the scaling behaviour only occurs at the percolation limit $p_c$ \cite{Blavatska2008}.
These results have been confirmed with numerical work \cite{Lee1988,Rintoul1994} and exact enumeration \cite{Lam1990,Ordemann2000,Nakanishi1991}. 
The addition of disorder also introduces new considerations such as how the type of averaging over disorder affects SAWs \cite{Nakanishi1992,Birkner2010} and when scaling laws are well-defined \cite{Janssen2007}.
In particular, we are interested in polymer collapse in a disordered medium.
Without disorder polymer collapse is a critical transition between a high-temperature extended phase and a low-temperature random globule phase known as the $\theta$ point.
It is also possible to have a third phase at low-temperature characterised that is collapsed but also more ordered than the globule phase and also maximally dense \cite{Bastolla1997}.
The canonical model for polymer collapse, the interacting self-avoiding walk (ISAW) can be extended to included stiffness, and this model exhibits a third phase characterised by anisotropic crystalline configurations and critical transitions to the extended and globule phases \cite{Krawczyk2009,Krawczyk2010}.
We previously \cite{Bradly2021} looked at the semi-stiff ISAW model on an inhomogeneous square lattice and found that the introduction of lattice defects causes a slight swelling of configurations in the globule phase and disrupts the formation of globally crystalline configurations in the crystal phase.
At larger amounts of inhomogeneity the critical transition between the globule and crystal phases disappears. 

In this work we look at another model for studying polymer collapse, using self-avoiding trails (SATs).
Whereas a SAW does not allow a lattice site to be visited more than once, a SAT relaxes this condition slightly, allowing sites to be visited more than once, but not bonds between sites.
Trails still exhibit the excluded-volume effect that makes these objects suitable for representing polymers, but can have slightly different properties to SAWs.
In particular, the question of whether collapse transitions in trail models are in the same universality class as for walk models  \cite{Owczarek1995, Prellberg1995, Owczarek2007}.
Polymer collapse with trail models works by assigning an interaction energy to sites with multiple visits.
By considering the trails on the triangular lattice we can assign different energies to doubly- or triply-visited sites, which induces another collapsed phase in two dimensions.
The homogeneous lattice case of this model has been studied previously \cite{Doukas2010}, showing that the collapse transition to the globule phase is $\theta$-like and the other collapsed phase is characterised by maximally dense configurations whose interior is dominated by triply-visited sites. The important difference to the third phase of the semi-stiff ISAW model is that this maximally dense phase is not ordered in a real crystalline sense as so may behave differently to the introduction of disorder. In another slightly different model \cite{Bedini2017} three collapsed phases were observed separately. To investigate the effect of disorder on this third type of collapsed phase we extend the model of  Doukas {\it et al.\ }\cite{Doukas2010} to include lattice inhomogeneity.

\section{Model and simulation}
\label{sec:Model}

We consider single polymers in dilute solution modelled as self-avoiding trails (SATs) on the triangular lattice.
The extended interacting SAT (eISAT) model allows for both doubly- and triply-visited sites, with different interactions energies based on the number of visits.
The canonical partition function for such SATs of length $n$ is
\begin{equation}
	Z_n(\omega_2, \omega_3) = \sum_{m_2,m_3} d_{n}(m_2,m_3) \, \omega_2^{m_2} \omega_3^{m_3},
	\label{eq:CombinedPartition}
\end{equation}
where $m_i$ is the number of sites with $i$ visits, $\omega_i$ is the Boltzmann weight for sites with $i$ visits and $d_{n}(m_2,m_3)$ is the density of states, or the number of configurations of length $n$, with $m_2$ doubly-visited sites and $m_3$ triply visited sites.
Here we consider both weights independently but certain special cases can be constructed by relating $\omega_3$ to $\omega_2$ \cite{Doukas2010,Owczarek1995}. 

\begin{figure}[t!]
	\centering
	\includegraphics[width=0.35\columnwidth]{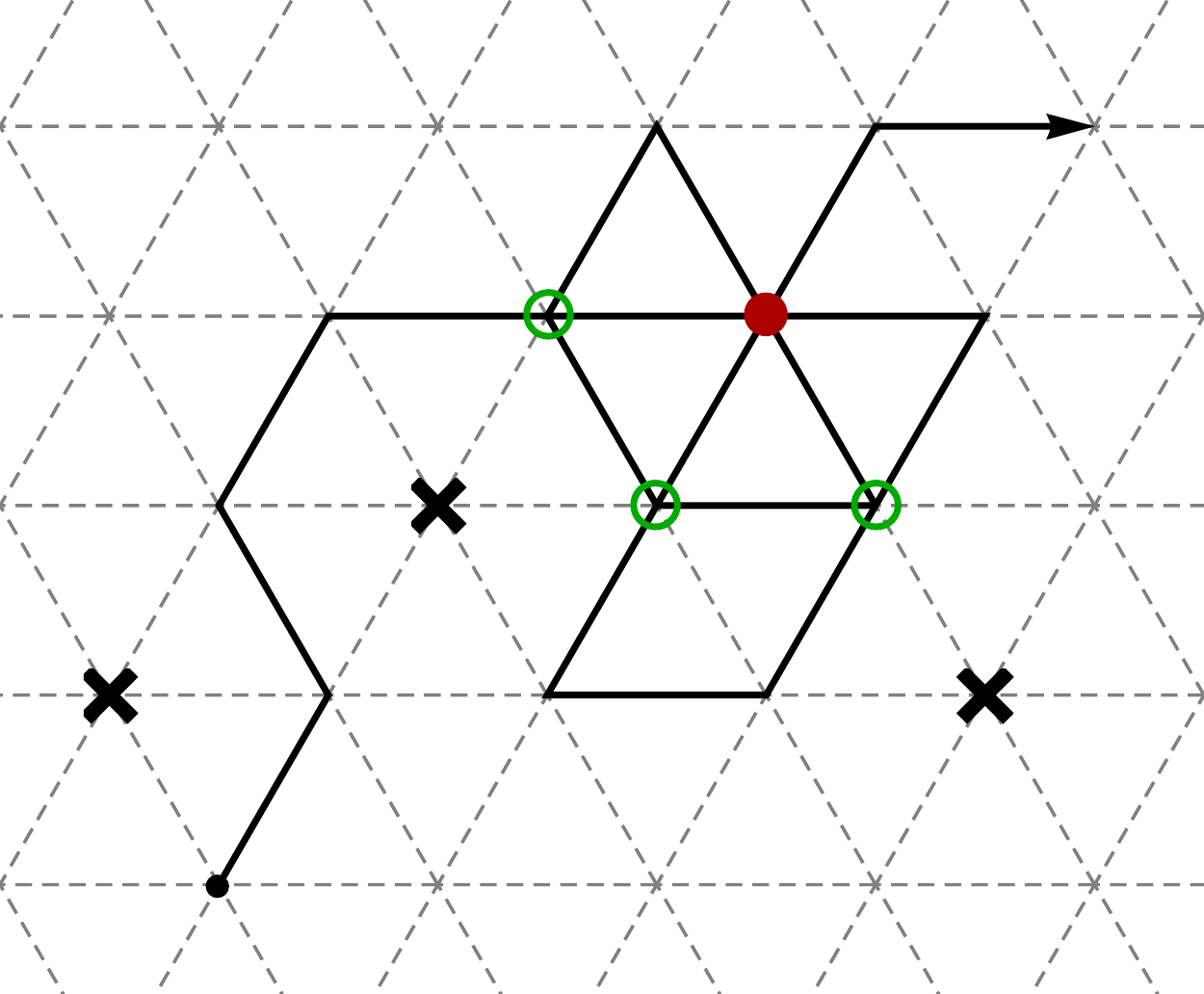}
	\caption{A self-avoiding trail on the triangular lattice with three doubly-visited sites (green circles) and one triply-visited site (red circle).
	Impurities in the lattice are marked with black crosses and prevent adjacent sites being triply-visited.}
	\label{fig:InhomoTriISAT}
\end{figure}

We represent the lattice defects like a site percolation model where lattice sites have a probability $p$ to be available.
This means a fraction $1-p$ of lattice sites is unavailable to the SAT and the partition function $Z_n(\omega_2, \omega_3; p)$ is now dependent on $p$.
We are interested in how the introduction of disorder affects the collapsed phases so we look at values of $1-p$ that are smaller than the percolation limit, which for site-percolation on the triangular lattice is $p_c = 1/2$ \cite{Stauffer1992}.
An example trail is shown in \fref{fig:InhomoTriISAT}.
Details of how the lattice configuration is chosen are given below when discussing the flatPERM algorithm.

To characterise the phases of the system we calculate the average density of doubly- and triply-visited sites $\avc$ and $\avd$, respectively.
For the transitions between these phases we consider the variance of parameter $m_i$,
\begin{equation}
	c_n^{(i)} = \frac{\text{var}(m_i)}{n} = \frac{\langle m_i^2 \rangle - \langle m_i \rangle^2}{n}.
\label{eq:VarM}
\end{equation}
In the thermodynamic limit this quantity becomes the specific heat which has singular behaviour $c_\infty \sim |T - T_c|^{-\alpha}$ governed by the universal scaling exponent $\alpha$.
If $\alpha < 1$ the transition is continuous and if $\alpha = 1$ then it is a first-order transition, in addition to a discontinuous jump in the densities.
For the finite-size system a crossover scaling ansatz is introduced and the singular part of the specific heat has the form 
\begin{equation}
	c_n \sim n^{\alpha\phi} \mathcal{F}\left[n^\phi(T-T_c)\right],
\label{eq:CnScalingModel}
\end{equation}
for some scaling function $\mathcal{F}$.
Near the critical point $T_c$ the scaling function is considered to be a positive constant and the exponent $\alpha$ can be found from the leading-order scaling of the peak of the variance 
\begin{equation}
	c_{n,\text{peak}}^{(i)} \sim n^{\alpha\phi}.
	\label{eq:CnPeakScaling}
\end{equation}
In some cases it is useful to consider the third derivative of the free energy $t_n$, whose peaks scale with exponent $(1+\alpha)\phi$.
Along with the well-known relation  $1/\phi = 2-\alpha$ \cite{Brak1993} the scaling of these quantities can be used to determine $\alpha$ and thus the nature of the transition.
For the full model it is useful to generalise the specific heat or variance to include the covariance of both parameters via the Hessian matrix
\begin{equation}
	H_n =
	\begin{pmatrix}
	 \frac{\partial^2 f_n}{\partial \omega_2^2} 	& \frac{\partial^2 f_n}{\partial \omega_2 \partial \omega_3}	\\
	 \frac{\partial^2 f_n}{\partial \omega_3 \partial \omega_2} 	& \frac{\partial^2 f_n}{\partial \omega_3^2}
	\end{pmatrix}
	,
	\label{eq:Hessian}
\end{equation}
where $f_n = -\tfrac{1}{n}\log Z_n$ is the reduced free energy.
The largest eigenvalue of $H_n$, which we denote $c_n^{(\lambda)}$, reduces to $c_n^{(i)}$ in cases where variance of one parameter $m_i$ is dominant.
In general, phase transitions are indicated by large $c_n^{(\lambda)}$.

In addition to derivatives of the free energy we are interested in metric quantities, for example the mean-square end-to-end distance
\begin{equation}
    \avR = \langle (\rr_{n} - \rr_0)^2 \rangle,
    \label{eq:EndToEndRadius}
\end{equation}
where $\rr_i$ is the position of the $i^\text{th}$ monomer in the chain.
The scaling of metric quantities is governed by the Flory exponent $\nu$, i.e.~$\avR \sim n^{2\nu}$.


The model is simulated using the flatPERM algorithm \cite{Prellberg2004}, an extension of the pruned and enriched Rosenbluth method (PERM) \cite{Grassberger1997}.
The simulation works by growing a trail up to some maximum length $N_\text{max}$ and counting the number of multiply-visited sites $m_2$ and $m_3$ at each step. Along the way the cumulative Rosenbluth \& Rosenbluth weight \cite{Rosenbluth1955} of the sample is recorded and used to update the sample weights $W_{n,m_2,m_3}$, which are an approximation to the athermal density of states $d_{n}(m_2,m_3)$ in \eref{eq:CombinedPartition}, for all $n\le N_\text{max}$.
FlatPERM prunes samples with low weight and enriches samples with high weight (relative to the current estimate of $W_{n,m_2,m_3}$) in order to maintain a flat histogram of samples over $n$, $m_2$, and $m_3$.
Flat histogram methods greatly enhance the sampling of low probability states, in this case those configurations with large values of $m_2$ and $m_3$.
The main output of the simulation are the weights $W_{n,m_2,m_3}$, from which thermodynamic quantities are calculated by specifying Boltzmann weights and using the weighted sum
\begin{equation}
    \langle Q \rangle_n(\omega_2,\omega_3) = \frac{\sum_{m_2,m_3} Q_{m_2,m_3} \omega_2^{m_2} \omega_3^{m_3} W_{n,m_2,m_3}}{\sum_{m_2,m_3} \omega_2^{m_2} \omega_3^{m_3} W_{n,m_2,m_3}}.
    \label{eq:FPQuantity}
\end{equation}

In certain cases it is advantageous to simulate a restricted model by fixing one of the Boltzmann weights $\omega_i$ at the beginning of the simulation.
The sum over the corresponding microcanonical parameter $m_i$ in \eref{eq:FPQuantity} is effectively performed within the simulation by altering the weight by a factor $\omega_i^{m_i}$.
The value of $m_i$ is only used locally at each step and the output weights array is two dimensional instead of three-dimensional for the full model.
The benefit is that the flatPERM algorithm is targeting a flat histogram in two parameters rather than three for the full model and so much larger lengths can be simulated in the same amount of time.
These restricted simulations correspond to a horizontal or vertical line in the $(\omega_2,\omega_3)$ parameter space which is useful for focusing on particular transitions in the phase diagram.

The inhomogeneous lattice is implemented by choosing a set of lattice sites to be inaccessible to the trail before it is grown.
The number of impurities is drawn from the appropriate binomial distribution with $p$ being the probability of any particular site being a valid site for the walk.
These impurities are distributed uniformly over the area of the lattice that would be accessible to a walk of length $n$.
The set of inaccessible sites is reseeded at the beginning of each flatPERM iteration (growing the walk from the origin).
The initial weight of each iteration is set to be the probability of the configuration of lattice impurities.
In this way the output weights $W_{n,m,s}$ contain the sum over disorder such that any $\langle Q \rangle$ in \eref{eq:FPQuantity} also represents a quenched-type average over disorder \cite{Nakanishi1992}.

It was recently demonstrated in \cite{Campbell2020} that a parallel implementation of the flatPERM algorithm is possible, whereby each thread grows samples independently but contributes to a global histogram and weights array in shared memory.
This is in contrast to the usual method of running multiple independent instances and then combining the results.
The shared memory approach does not simulate samples at a higher rate but does have the advantage that the approach to equilibrium is much faster in the early stages of running, and thus the algorithm does not need to be run for as long to achieve similar results to the serial implementation.
In this work we employ a parallel implementation of flatPERM to simulate both the restricted and full eISAT models, which involve two and three microcanonical parameters, respectively.
We still run several independent simulations for the same model with each independent instance using multiple threads in parallel.
This provides a measure of statistical uncertainty as well as enough iterations to properly sample the lattice defect configurations.
We thus effectively employ 100s of CPU threads for each model enabling us to simulate $10^5$ iterations of the full model up to length $n = 600$ and $10^6$ iterations of the restricted model up to length $n = 1444$ in less than 100 hours of server walltime, compared to smaller lengths taking several weeks with a serial implementation.
These system sizes are significantly greater than earlier studies of the eISAT model \cite{Doukas2010} and the semi-stiff ISAW model on the inhomogeneous lattice \cite{Bradly2021}.
We are also aided by the fact that self-avoiding trails are sampled slightly faster than self-avoiding walks, since trails typically have more moves available at each step so less pruning is required.
We also remark that our implementation ignores race conditions from a shared memory implementation but that this has little or no effect on efficiency for the system sizes considered.
This is similar to naive parallelisation that can be applied to the Wang-Landau algorithm \cite{Zhan2008}.

\begin{figure}[ht!]
\centering
\includegraphics[width=\textwidth]{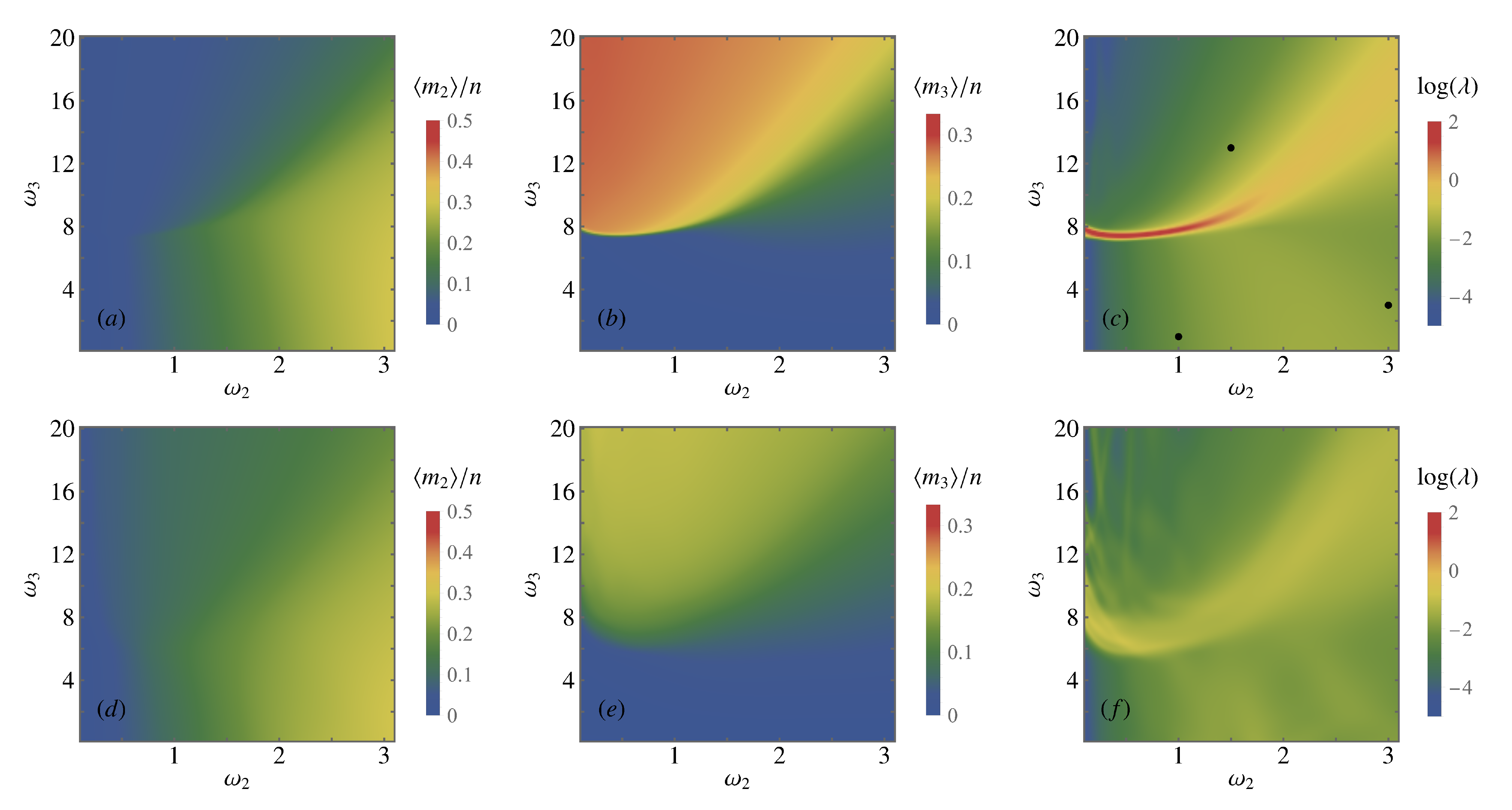}
\caption{The behaviour of the model in the full phase space is elucidated by considering the average densities of doubly-visited sites $\avc$ (left) and triply-visited sites $\avd$ (middle) and the logarithm of the largest eigenvalue of the covariance matrix $H_n$ (right). In this way a phase diagram can be inferred.
Plots are for length $n = 600$ and $1-p = 0$ (top) and $1-p = 0.2$ (bottom).
Black points in (c) refer to typical configurations of \fref{fig:Configurations}.}%
\label{fig:FullPhase}%
\vspace{0.25cm}
\includegraphics[width=\textwidth]{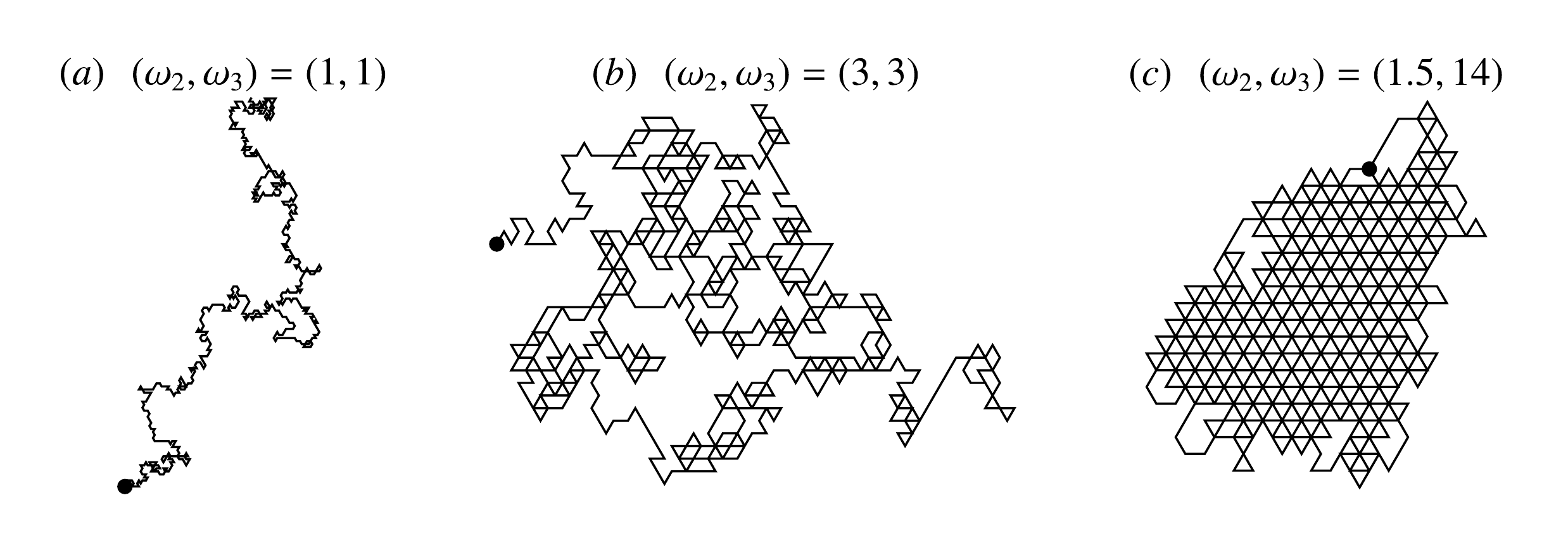}
\includegraphics[width=\textwidth]{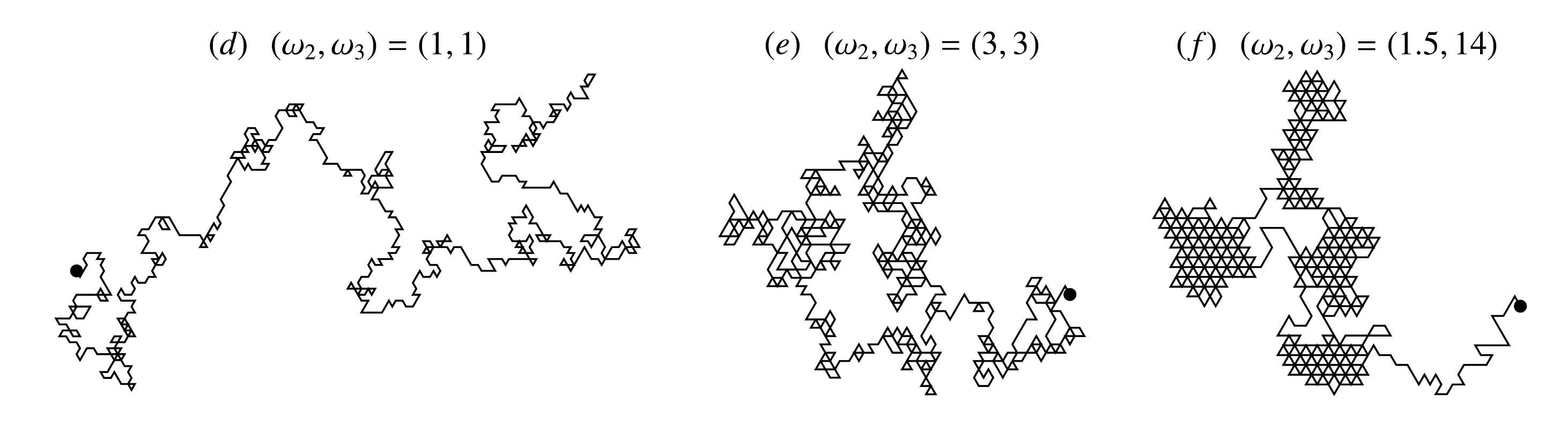}
\caption{Typical configurations at points in the phase space indicated on \fref{fig:FullPhase}(c), which corresponds to the swollen phase, globule phase and maximally dense phase, respectively. Top row (a-c) are for the homogeneous lattice with $1-p = 0$ and bottom row (d-f) are for the inhomogeneous lattice $1-p = 0.2$.}%
\label{fig:Configurations}%
\end{figure}

\section{Phase diagram}
\label{sec:Phase}

First we characterise the phases by looking at the densities and the expected configurations, with and without lattice impurities.
For this we simulated the full eISAT model up to maximum length $n = 600$ using parallel flatPERM.
In \fref{fig:FullPhase} we plot the average density of doubly-visited sites $\avc$ (left) and average density of triply-visited sites $\avd$ (middle).
The variance of the microcanonical parameters is also shown in the plots on the right, which plot the logarithm of the largest eigenvalue $\lambda$ of the covariance matrix $H_n$, \eref{eq:Hessian}.
The top row is for the homogeneous lattice, $1-p = 0$, and the bottom row is with lattice defects present, $1-p = 0.2$.
Further visualisation of the phases is given in \fref{fig:Configurations} which shows typical configurations at points in the $(\omega_2,\omega_3)$ phase diagram that are indicative of each phase.
These points are marked with black dots on \fref{fig:FullPhase}(c).

On the homogeneous lattice, $1-p = 0$ we infer that there are three phases, as previously conjectured \cite{Doukas2010}.
For small $\omega_2$ and $\omega_3 \lesssim 8$ the extended phase is characterised by both densities $\avc$ and $\avd$ being very small, though non-zero. In this phase the trails are in an extended or swollen configuration like in \fref{fig:Configurations}(a). We confirm below that the Flory exponent is the expected $\nu=3/4$.
For larger $\omega_2$ the system enters the globule phase characterised by collapsed configurations as in \fref{fig:Configurations}(b).
Here $\avc$ has a significantly larger value that smoothly increases as $\omega_2$ increases, trending to the maximum value $1/2$ at very large $\omega_2$ (very low temperature).
The density of triply-visited sites, $\avd$, is still small for $\omega_3 \lesssim 8$, but starts to increase as $\omega_3 $ increases, which we argue below is the approach to a maximally dense phase.
The transition to the globule phase from the extended phase is expected to be $\theta$-like and occurs at a critical value $\omega_2^\text{c}$ that depends on $\omega_3$ and decreases as $\omega_3$ increases.
However, it is a weak transition and it is difficult to make out even on the logarithmic scale of \fref{fig:FullPhase}(c).
Lastly, the maximally dense phase appears for large $\omega_3$ where $\avc$ again becomes small and would vanish as $\avd$ quickly approaches its maximum value of $1/3$. In fact, the phase is expected to be characterised by the thermodynamic limit $\lim_{n\rightarrow \infty} \avd = 1/3$ for any point $(\omega_2,\omega_3)$ in this phase. 
\fref{fig:Configurations}(d) shows a typical configuration in this phase were the trail is dense in the interior with only a small fraction of the trail in singly- or doubly-visited sites, mainly on the boundary. 
The transition to the maximally dense phase from the extended phase is first-order, shown by a line of high variance in \fref{fig:FullPhase}(c).
The transition from the globule phase to the maximally dense phase is continuous, but appears stronger than the $\theta$-like extended-globule transition.
It is expected that the phase boundaries meet at the multi-critical point $(\omega_2,\omega_3) = (5/3,25/3)$, where the eISAT model corresponds to unweighted pure  kinetic growth of trails \cite{Doukas2010}.
In our finite size data where the phase boundaries meet differs from the exact kinetic growth point by a small but noticeable amount, despite the much longer length we simulate here, suggesting that there are still sizable finite-size corrections to consider.

In the case of the inhomogeneous lattice, $1-p = 0.2$, where a considerable fraction of the lattice is unavailable to the walks, the extended and globule phases are largely unchanged but there are several differences regarding the maximally dense phase.
The extended phase is still characterised by small values of $\avc$ and $\avd$.
In the globule phase $\avd$ is very close to zero, except near the transition to the maximally dense phase, and $\avc$ has a larger finite value increasing with $\omega_2$ though still small compared to its possible maximum of $1/2$.
The configurations, shown in \fref{fig:Configurations}(e,f), have the same character as the homogeneous lattice.
The transition between the extended and globule phases is still too weak to seen on this scale, even when the other transitions are weakened by the presence of lattice defects.
The largest change when lattice inhomogeneity is introduced is the disruption to the maximally dense phase.
Firstly, the densities have significantly different values compared to the homogeneous lattice case.
Comparing \fref{fig:FullPhase}(a) and (d), we see that the density of doubly-visited sites $\avc$ is now non-zero for $\omega_2 > 1$ and large $\omega_3$.
From \fref{fig:FullPhase}(b) and (e) we also see that the density of triply-visited sites $\avd$ is reduced but still substantial.
When looking at the variances in \fref{fig:FullPhase}(c) and (f) it appears that the sharp first-order transition boundary between the extended and maximally dense phases is gone.
There is evidence that a weaker transition remains in roughly the same place and, in fact, the same could be  the case for the globule-maximally dense transition. However, if the maximally dense phase disappears and becomes simply a denser version of the globule phase there can be no thermodynamically sharp transition. 
The finite size nature of this analysis urges caution and a conservative interpretation suggests that there is a smooth transition as $\omega_3$ is increased for large $\omega_2$.  
In fact there are many artefacts arising from the difficulty to obtain good convergence for low temperatures in \fref{fig:FullPhase}(f) that make it difficult to ascertain the phase diagram clearly and we will look more closely at some of these possible transitions below. 
Lastly, we note that the kinetic growth model does not map to a critical point of the ISAT model on the inhomogeneous lattice because the presence of defects allows for the kinetic growth trails to become trapped and it is also worth noting that a mapping of kinetic growth to a static model induces an interaction with the defect.

From these plots of the densities, it appears at first sight that there is a difference between the eISAT model and the semi-stiff ISAW model \cite{Bradly2021} regarding the effect of the lattice inhomogeneity on the maximally dense and crystal phases.
In the latter case, lattice inhomogeneity clearly erased the distinction between the globule and crystal phases as the lattice defects prevented anisotropic configurations and the phase diagram showed only a extended phase and a collapsed phase (\cite{Bradly2021} Fig.~2).
In the eISAT model there still seems to be a transition between the globule phase and the region of the phase diagram that contained the maximally dense phase on the homogeneous case in respect that the densities even if the difference is smaller.
Regarding the typical configurations, \fref{fig:Configurations}(h) shows that the lattice inhomogeneity breaks the trail into several sub-clusters, each exhibiting a maximally dense interior. However the overall configuration is no longer maximally dense.
The separation into clusters (blobs) joined by strands of singly-visited sites, and thus an increase in the size of the surface relative to the bulk, accounts for the increase in $\avc$ and the decrease in $\avd$ compared to the homogeneous lattice case. So from this point of view the maximally dense phase is replaced by a denser version of the globule phase where the blobs become dense.
This is similar to the semi-stiff ISAW model where well separated sub-clusters form, each with internal anisotropy.
However, the subtle difference is that in that model the global anisotropy of the whole walk becomes drastically reduced when lattice inhomogeneity is introduced since the sub-clusters are not correlated. Overall, this reinforces our interpretation that the maximally dense phase is broken and no real transition between small and large $\omega_3$ occurs.

\begin{figure}[t!]
\centering
\includegraphics[width=0.7\columnwidth]{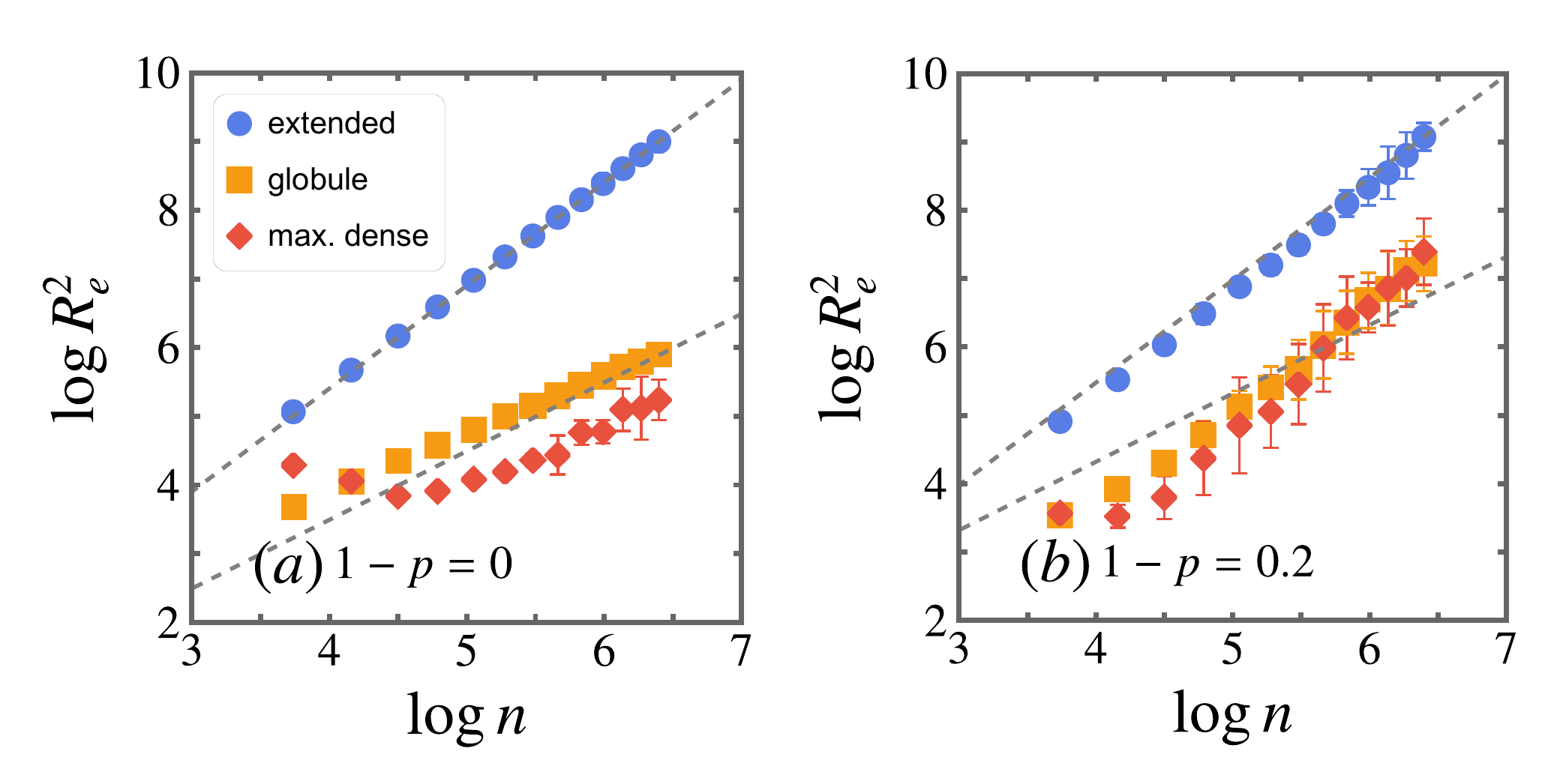}
\caption{The mean-squared end-to-end distance $\avR$ without and with lattice inhomogeneity at representative points of each phase.
Data is from the full model up to length $n = 600$.
Dashed reference lines indicate scaling corresponding to $\nu = 1/2, 3/4$.}%
\label{fig:R2Scaling}%
\end{figure}

The prime issue is of finite size scaling and the effective lengths at which our simulations are performed. One way to understand this is via the scaling of metric quantities, for example the mean-square end-to-end distance $R_n^2 \sim n^{2\nu}$.
In two dimensions the exponent has well-known values $\nu = 3/4$ in the extended phase, and $\nu = 1/2$ in collapsed phases.
In \fref{fig:R2Scaling} we show log-log plots of $R_n^2$ at points in the phase diagram representing each of the three phases.
Although the specific values of the weights do not matter for this picture, the data for each phase is: extended, $(\omega_2,\omega_3) = (1,1)$; globule, $(\omega_2,\omega_3) = (5,1)$; and maximally dense, $(\omega_2,\omega_3) = (1,20)$.
On the homogeneous lattice (a) all phases have expected scaling. 
Note that the maximally dense phase is not well-formed for the smallest values of $n$ even at the large value of $\omega_3$ chosen as the representative point and so the data for this phase does not indicate any real scaling behaviour until larger $n$.
On the inhomogeneous lattice (c) with $1-p = 0.2$, the scaling in the collapsed phases clearly departs from $\nu = 1/2$ at all values of $n$ and $\avR$ appears to scale with an effective finite size exponent between $\nu=1/2$ and $\nu = 3/4$. This indicates that the lengths of our simulations are too small to proper see the low temperature behaviour in a finite size scaling analysis. The alternate explanation is that the impurities not only disrupt the maximally dense phase but also destroy the globular phase. This was not seen for the ISAW model but the lengths of those simulations were shorter than we have conducted here. We shall return to this point in the conclusion. Important for this work is that  in the presence of impurities the trails appear to behave in the same way in both collapsed regions of the phase space.

\section{Phase transitions}
\label{sec:PhaseTransitions}
We now consider the each of the homogeneous phase transitions and how they are affected by the introduction of defects more closely as the amount of defects becomes small.
\subsection{Extended-globule transition}
\label{sec:ExtendedGlobuleTransition}

\begin{figure}[t!]
\centering
\includegraphics[width=0.7\columnwidth]{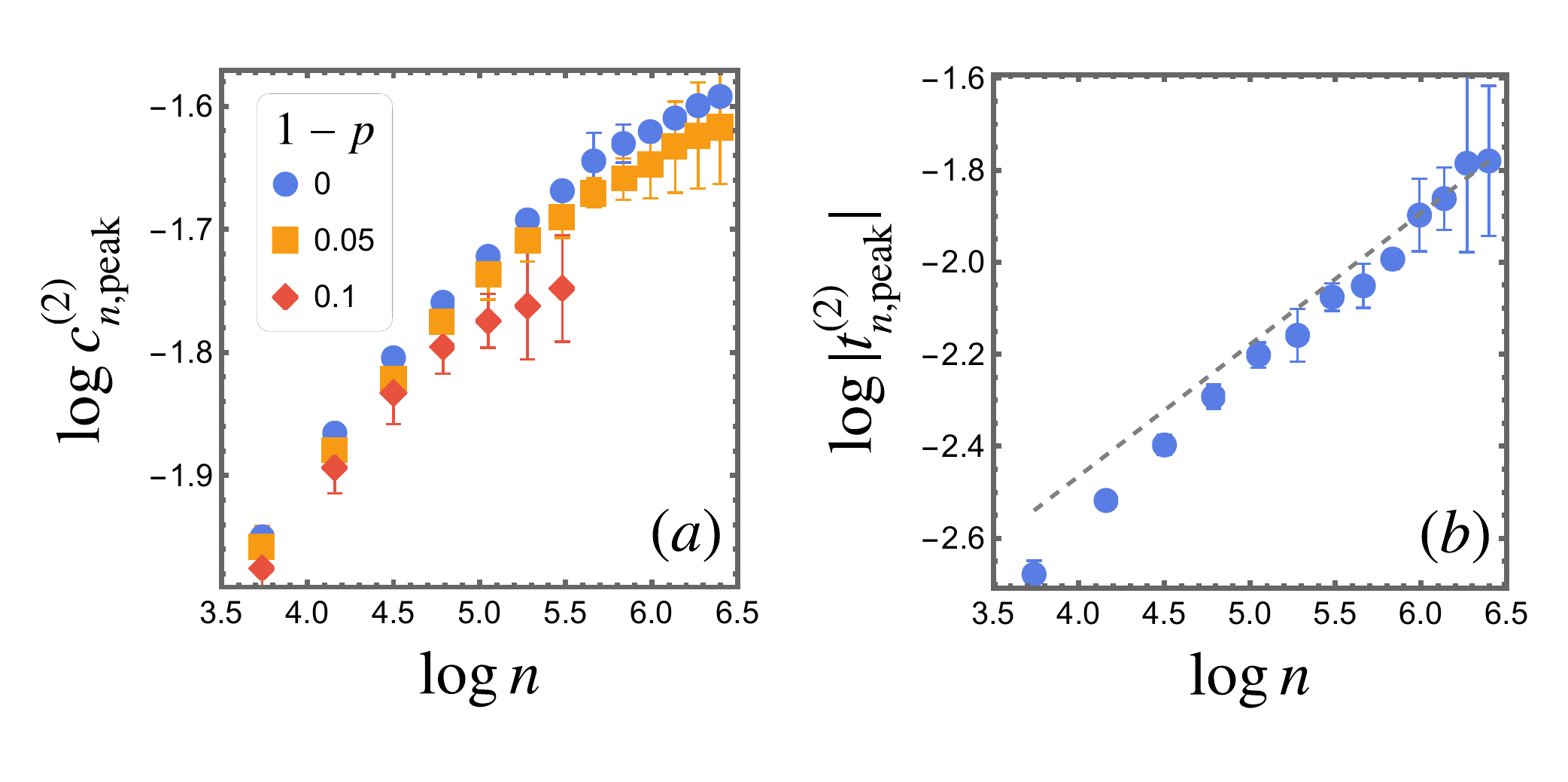}
\caption{The peak values of (a) variance of doubly-visited sites $c_n^{(2)}$ and (b) the third derivative of the free energy $t_n^{(2)}$ near the extended-globule transition for $\omega_3 = 5$. 
Data is from the full model simulations up to length $n = 600$.}
\label{fig:CnPeaksExtGlob}%
\end{figure}

We first look at the critical transition between the extended and globule phases.
As we have seen in \fref{fig:FullPhase}, this transition is weaker than the others and on the homogeneous lattice it is expected to be a $\theta$-like transition.
In two dimensions the $\theta$ point transition is characterised by $\alpha = -1/3$, thus the peak value of the variance $c_n^{(2)}$ does not diverge and the scaling form of \eref{eq:CnPeakScaling} is not useful.
However, the peak of the third derivative of the free energy $t_n^{(2)}$ \emph{does} diverge, with exponent $2/7$, and we can visualise the peak values to determine the nature of the transition.
We consider moments of $m_2$ as the indicators of this transition, since $\avd$ changes only slowly near this transition.
In \fref{fig:CnPeaksExtGlob} we plot the peak values of (a) the variance $c_n^{(2)}$ and (b) the third derivative of the free energy $t_n^{(2)}$ using data from the full model but at a fixed value $\omega_3 = 5$, across the extended-globule transition.
For both the homogeneous lattice and and inhomogeneous lattice with small amount of defects, $1-p = 0.05$, the peaks in $c_n^{(2)}$ are clear.
For larger amount of inhomogeneity, the peaks are only clear for a smaller range in $n$; for larger lengths the peaks are indistinguishable from the numerical noise.
Where the peaks are well-defined, their magnitudes diverge slowly with increasing $n$ and corrections to scaling are significant, judging by the curvature of the data.
For the homogeneous case we can show in (b) that the peaks of $t_n^{(2)}$ for the homogeneous lattice do diverge, along with a dashed line with slope $2/7$.
Thus, we see that the extended-globule transition on the homogeneous lattice has the expected $\theta$-like characteristics.
The data for the inhomogeneous lattice cases is inconclusive on this point due to significant noise in the data.
The extended-globule transition persists on the inhomogeneous lattice, at least for small values of $1-p$, but we cannot be definitive about the nature of this transition,
although it is expected to remain a $\theta$-like transition \cite{Duplantier1988}.

\subsection{Globule-maximally dense transition}
\label{sec:GlobuleDenseTransition}

\begin{figure}[t!]
\centering
	\includegraphics[width=0.7\columnwidth]{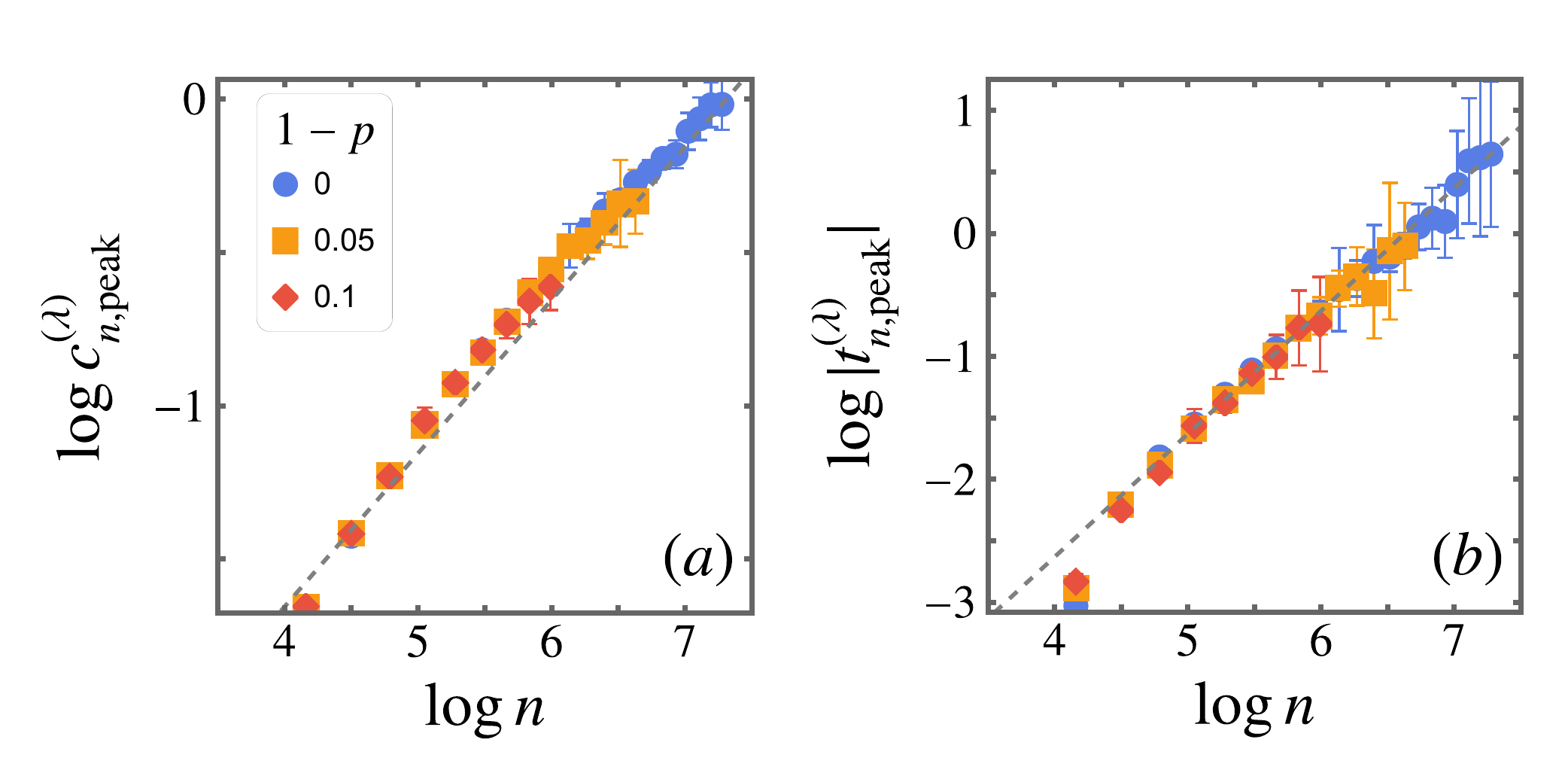}
\caption{The peak values of (a) $c_n^{(\lambda)}$ and (b) $|t_n^{(\lambda)}|$ near the globule-maximally dense phase transition for several amounts of lattice inhomogeneity. 
Data is from restricted model simulations at $\omega_2 = 3$ up to length $n = 1444$.
Reference lines a show scaling for exponent $\alpha = 1/2$.}%
\label{fig:CnPeaksGlobMax}%
\end{figure}

Next, we consider the transition between the globule and maximally dense phases.
We ran additional simulations of the restricted model with fixed $\omega_2 = 3$ up to length $n = 1444$.
Since both phases are collapsed we look at the covariance $c_n^{(\lambda)}$ for a signature of a transition.
In \fref{fig:CnPeaksGlobMax} we show (a) a log-log plot of peaks of $c_n^{(\lambda)}$ and (b) a log-log plot of the peaks of $|t_n^{(\lambda)}|$.
In the homogeneous lattice case we expect a continuous transition with scaling exponent close to $\alpha = 1/2$.
Although we do not have enough data to estimate $\alpha$ or corrections to scaling accurately, the data appears consistent with this exponent, shown by the reference lines in the plots.

For the inhomogeneous lattice cases we only plot points for a limited range of $n$ where the peaks are distinct.
At larger $n$ there is not a clear peak indicating a transition and this valid range shrinks as $1-p$ decreases.
Within this valid range the magnitudes of the peaks of $c_n^{(\lambda)}$ overlap well with the homogeneous lattice case.
This suggests that for lengths that are not too disturbed by the lattice defects the transition exists and is unaltered.
This behaviour persists until some maximum length, dependent on $1-p$, after which the transition is not evident and the two collapsed phases merge.

\subsection{Extended-maximally dense transition}
\label{sec:ExtendedDenseTransition}

\begin{figure}[t!]
	\centering
	\includegraphics[width=0.35\linewidth]{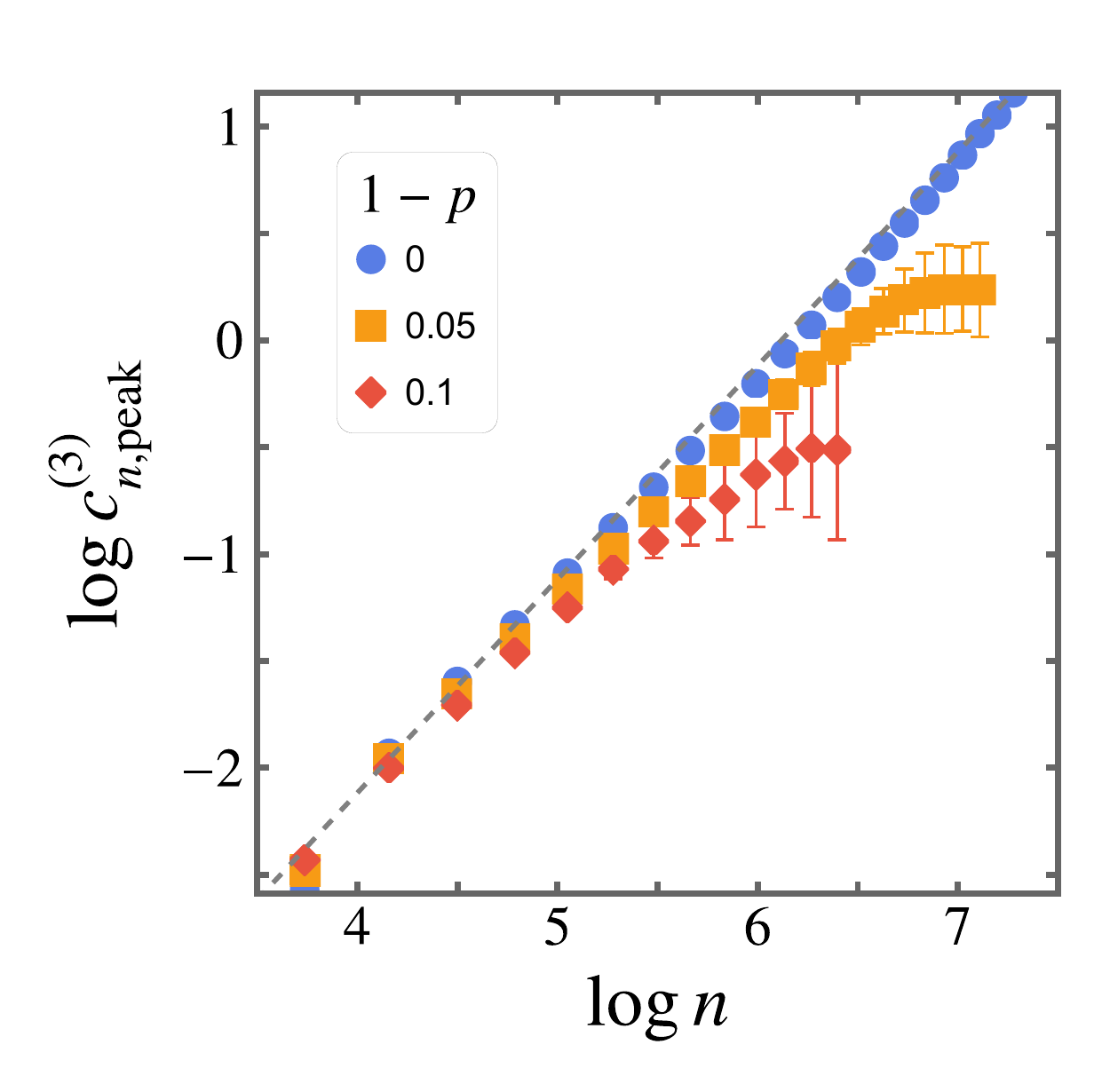}
\caption{The peak values of the variance of triply-visited sites $c_n^{(3)}$ near the extended-maximally dense phase transition for several amounts of lattice inhomogeneity. 
The dashed reference line has a slope of 1.
Data is from restricted model simulations with fixed $\omega_2 = 1.5$ up to length $n = 1444$.}%
\label{fig:CnPeaksExtMax}%
\end{figure}

To look at the extended-maximally dense transition more closely we ran additional simulations of the restricted model with fixed $\omega_2 = 1.5$ up to length $n = 1444$.
In \fref{fig:CnPeaksExtMax} we plot the peaks of the variance of triply-visited sites $c_{n,\text{peak}}^{(3)}$ near the extended-maximally dense transition.
In the case of the homogeneous lattice $1-p = 0$, the first order nature of the transition is clear, since the peaks scale linearly with $n$ suggesting an exponent $\alpha = 1$.
In the presence of a small amount of inhomogeneity, $1-p = 0.05$, the linear scaling persists up to some maximum, and this maximum reduces as inhomogeneity increases to $1-p = 0.10$.
Similar to the globule-maximally dense transition, at large $n$ the variance has no identifiable peak to indicate a transition and these points are not shown on \fref{fig:CnPeaksExtMax}.
Unlike the globule-maximally dense transition however, there is a small window where a peak can be identified but the magnitude has sublinear scaling. So we can confidentially conclude that the first order transition disappears but less confident about its replacement.
If the addition of a small amount of lattice inhomogeneity allows a single collapsed phase to persist but without a distinction between globule and maximally dense phases then one expects that the extended-maximally dense transition must change to match the extended-globule transition, which we know to be at least continuous, possibly $\theta$-like.
The fact that there is a small window in the data where this may occur is tantalising but but we cannot be conclusive.
We do not have reliable enough data to probe with certainty, for example even where peaks in the variance can be identified, the simulations needs further convergence to reliably estimate the third derivative $t_n$ and thus the continuous transition scaling.

\section{Crossover to disordered system}
\label{sec:CrossoverToDisorder}

\begin{figure}[t!]
\centering
\includegraphics[width=0.7\columnwidth]{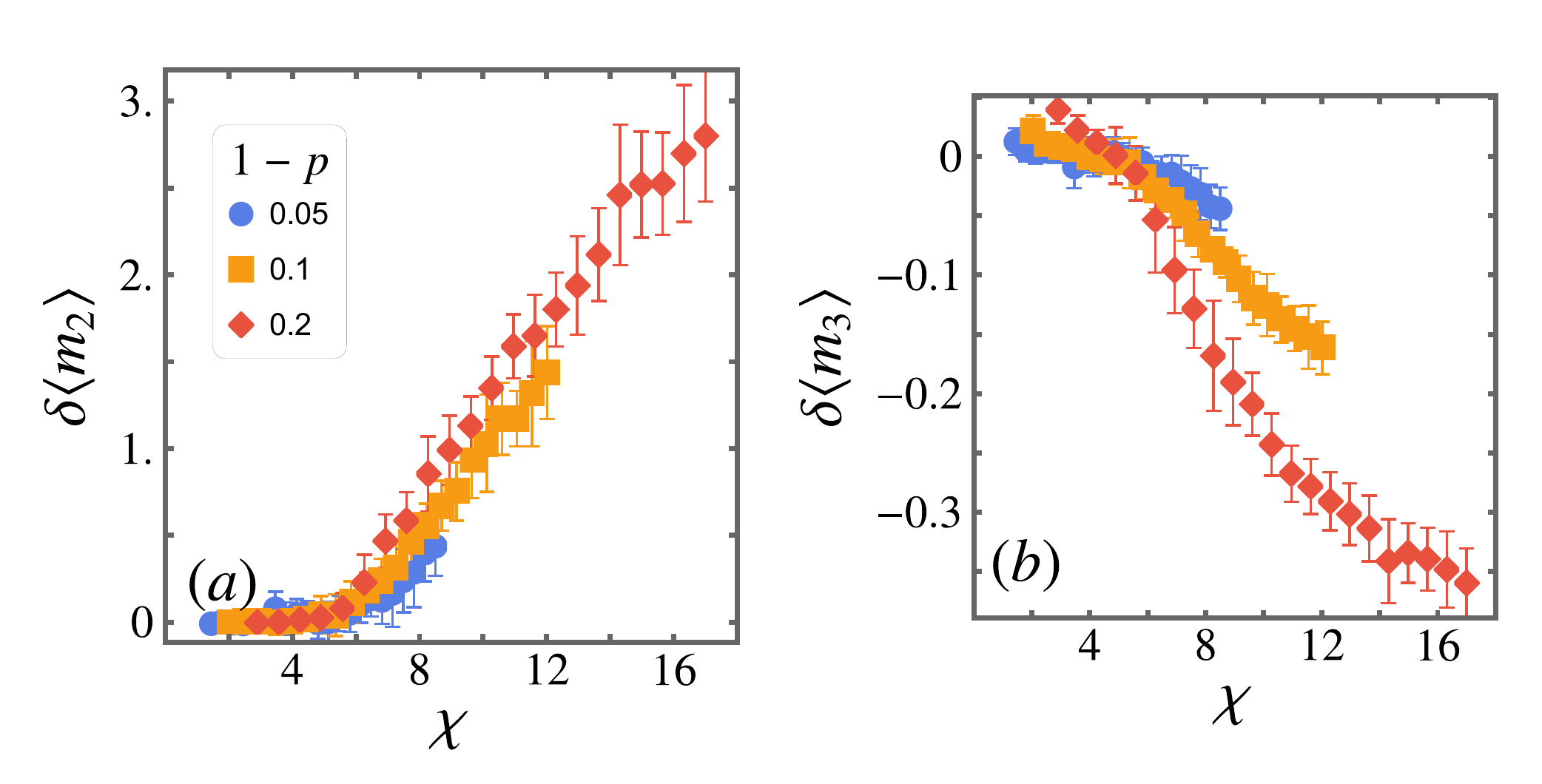}
\caption{The densities of the inhomogeneous lattice model relative to the homogeneous lattice model, versus the scaling parameter $\chi$, in the maximally dense phase, $(\omega_2,\omega_3) = (3,100).$}%
\label{fig:MiVsChi}%
\end{figure}

The extent of the disruption caused by increasing inhomogeneity is different for each transition and each phase.
However, a common feature is that as the inhomogeneity increases, there is a range in $n$ where expected behaviour persists, and above these lengths the transitions are altered to some degree.
The more inhomogeneity is present, the smaller this range is but it is somewhat \emph{ad hoc} to determine this range from where the scaling behaviour of $c_{n\text{peak}^{(i)}}$ changes.
Since we have a finite-size system the obvious way to characterise the amount of disorder is by the parameter $\chi = n^\nu\sqrt{1-p}$, which is the ratio of the leading order scaling of metric quantities (e.g.~end-to-end distance) to the mean separation of defects $1/\sqrt{1-p}$.
We are focused on the maximally dense phase so we will use the collapsed phase value for the exponent $\nu$, i.e.~$\nu = 1/2$ and see where this breaks down.
As a measure of the effect of the lattice defects we look at the densities in the inhomogeneous lattice cases relative to the homogeneous lattice case.
These quantities have smaller numerical uncertainty from simulations on an inhomogeneous lattice, compared to the variances consider in the previous section.
We define
\begin{equation}
	\delta \langle m_i \rangle = \frac{\langle m_i \rangle_p - \langle m_i \rangle_0}{\langle m_i \rangle_0},
	\label{eq:DeltaMi}
\end{equation}
where $\langle m_i \rangle_p$ and $\langle m_i \rangle_0$ are the densities calculated for the inhomogeneous and homogeneous lattice cases, respectively.
In \fref{fig:MiVsChi} we plot $\delta \langle m_i \rangle$ as a function of $\chi$ using data from the restricted model with fixed $\omega_2 = 3$ at a large value of $\omega_3 = 100$ to highlight the effect in the maximally dense phase.
It is worth remarking that at this point in the phase diagram $\langle m_2 \rangle_p$ is small and $\langle m_3 \rangle_p$ is close to $1/3$, regardless of $1-p$.

We identify low- and high-disorder regimes, delineated around $\chi \approx 6$.
This point is common to both densities and it also corresponds to the values of $n$ where the peaks of the variances change behaviour in \sref{sec:PhaseTransitions}.
In \fref{fig:MiVsChi}(a) $\delta \langle m_2 \rangle$ is largely independent of the inhomogeneity in the low-disorder regime, where lattice defects are present but are too few to disrupt very dense configurations.
There is a marked change in behaviour in the high-disorder regime where $\delta \langle m_2 \rangle$ increases with $\chi$; there is still some small dependence on $1-p$ but it is not clear from this data if this is significant.
There are two possible effects that contribute to this enhancement.
Firstly, a lattice defect prevents triply-visited sites in its immediate vicinity so more doubly-visited sites appear in the interior of a configuration.
Secondly, lattice defects inhibit a single dense globule in favour of more smaller sub-clusters thus increasing the surface of the configuration (where doubly-visited sites appear) relative to the bulk (dominated by triply-visited sites).
Judging by the most probable configurations shown in \fref{fig:Configurations} it seems that the second effect is stronger.

The effect of inhomogeneity on the density of triply visited sites is different, shown in \fref{fig:MiVsChi}(b).
In the low-disorder regime $\delta \langle m_3 \rangle$ appears to be enhanced relative to the homogeneous lattice case, but this is actually a finite-size effect as the enhancement decreases as $\omega_3$ is increased.
We speculate that a small amount of inhomogeneity inhibits the average size of configurations which reduces the size of the surface (dominated by doubly-visited sites) relative to the bulk (dominated by triply-visited sites).
Recall that for larger $1-p$ small $\chi$ corresponds to smaller $n$, where this effect is more significant.
In the high-disorder regime $\delta \langle m_3 \rangle$ is reduced as $\chi$ increases and a residual dependence on $1-p$ is more prominent.

\begin{figure}[t!]
\centering
\includegraphics[width=0.7\columnwidth]{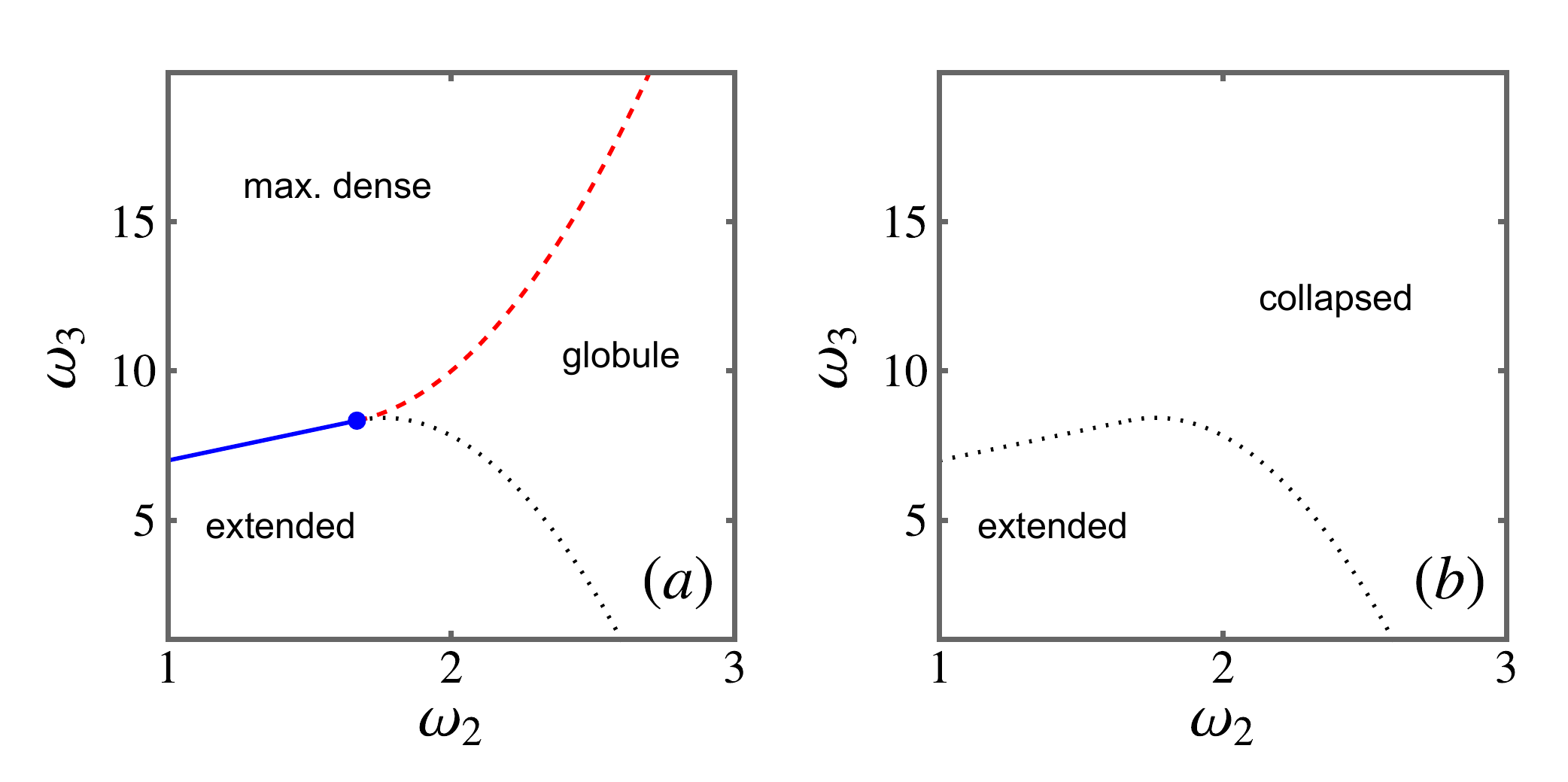}
\caption{Schematic phase diagrams for (a) low disorder, including homogeneous lattice, and (b) near $\chi \approx 6$.
The solid blue line is a first-order transition, the dotted black line is a $\theta$-like phase transition and the dashed red line is a continuous phase transition.}%
\label{fig:SchematicPhases}%
\end{figure}

We summarise our findings in \fref{fig:SchematicPhases} with two schematic phase diagrams.
When the amount of disorder is zero or asymptotically small there is a scaling regime, $\chi \lesssim 6$, which includes the homogeneous lattice case, such that the system contains three phases, shown in (a). 
In this phase diagram the behaviour of the transitions between the phases is known including that they meet at a multi-critical point.
Thermodynamically then this phase diagram is only valid for the homogeneous lattice but there is a scaling regime characterised by $\chi$. 
At some point around $\chi \approx 6$ the maximally dense phase is disrupted and the transition to the globule phase disappears. 
Further, the extended-maximally dense transition changes to a continuous one and in order to be consistent with what was the extended-globule transition, we expect that it becomes $\theta$-like.
We show in (b) a schematic phase diagram for finite impurity case with only two phases. 
It is possible that the phase boundaries may have shifted relative to the small $\chi$ phase diagram, but we cannot quantify this shift.
However, we do expect that the phase boundary, if it does exist in this regime, does not include the kinetic growth point from the homogeneous lattice case.
Of course, the alternate hypothesis is that there are no longer any phase boundaries for fixed finite levels of impurities in the thermodynamic limit. 
The resolution of this question requires further work with longer length simulations.

\section{Conclusion}
\label{sec:Conclusion}

We have simulated the extended ISAT model of lattice polymers on the homogeneous and inhomogeneous triangular lattices.
The presence of lattice defects disrupts the maximally dense phase and the transitions to the extended and globule phases in different ways.
This work complements a previous study of the semi-stiff ISAW model on the square lattice \cite{Bradly2021}.
In that model the low temperature analogue to the maximally dense phase is a crystal phase (also maximally dense but with added anisotropy) characterised by closely packed long straight segments.
It was intuitive that lattice defects would inhibit such crystalline configurations and this was most apparent in the average anisotropy of the configurations.
In particular, the value of the anisotropy in the crystal phase displayed crossover behaviour between low and high disorder regimes when parameterised by an appropriate scaling parameter. 
Anisotropy is not useful in the eISAT model but by introducing the same scaling parameter $\chi$ we find a crossover scaling between homogeneous and inhomogeneous lattice regimes.
The crossover is apparent in the values of the densities $\avc$ and $\avd$ in the maximally dense phase and the scaling of peaks of $c_n$ near the transitions.
Although the maximally dense phases in the eISAT model is different to the crystalline phase in the ISAW model, the introduction of lattice defects disrupts these dense phases in similar ways, causing the formation of dense sub-clusters.
Our findings are consistent with the expectation that a critical transition between the extended and collapsed phases persists as the amount of lattice inhomogeneity increases and that the transition between the globule and maximally dense phase becomes a thermodynamically smooth change. However, our simulations sizes are not large enough to verify exponents. One question that needs addressing with longer simulations is whether lattice impurities also disrupt the globule phase.

\begin{acknowledgements}
Financial support from the Australian Research Council via its Discovery Projects scheme (DP160103562) is gratefully acknowledged by the authors. 
\end{acknowledgements}

\bibliography{../../Manuscripts/polymers_master}{}

\end{document}